# Agentic AI for SAGIN Resource Management: Semantic Awareness, Orchestration, and Optimization

Linghao Zhang, Haitao Zhao, Bo Xu, Hongbo Zhu, and Xianbin Wang, *Fellow, IEEE*

*Space-air-ground integrated networks (SAGIN) promise ubiquitous 6G connectivity but face significant resource management challenges due to heterogeneous infrastructure, dynamic topologies, and stringent quality-of-service (QoS) requirements. Conventional model-driven approaches struggle with scalability and adaptability in such complex environments. This paper presents an agentic artificial intelligence (AI) framework for autonomous SAGIN resource management by embedding large language model (LLM)-based agents into a Monitor-Analyze-Plan-Execute-Knowledge (MAPE-K) control plane. The framework incorporates three specialized agents, namely semantic resource perceivers, intent-driven orchestrators, and adaptive learners, that collaborate through natural language reasoning to bridge the gap between operator intents and network execution. A key innovation is the hierarchical agent-reinforcement learning (RL) collaboration mechanism, wherein LLM-based orchestrators dynamically shape reward functions for RL agents based on semantic network conditions. Validation through UAV-assisted AIGC service orchestration in energy-constrained scenarios demonstrates that LLM-driven reward shaping achieves 14% energy reduction and the lowest average service latency among all compared methods. This agentic paradigm offers a scalable pathway toward adaptive, AI-native 6G networks, capable of autonomously interpreting intents and adapting to dynamic environments.*

## Introduction

Space-air-ground integrated networks (SAGIN) are emerging as a critical infrastructure for 6G and beyond, promising seamless global connectivity and ubiquitous service delivery across terrestrial, aerial, and space users. By synergistically integrating satellites, aerial platforms, and ground infrastructure, SAGIN can support diverse applications such as emergency response, remote sensing, and intelligent transportation systems [1]. However, realizing this vision faces fundamental challenges arising from the stark heterogeneity and strong coupling of network segments. SAGIN inherently integrates resource-constrained satellite networks, which are characterized by broad coverage despite limited onboard energy and computing capabilities, with resource-rich terrestrial networks. This integration creates a complex environment exhibiting drastic performance heterogeneity in terms of latency, data rates, and energy efficiency. Furthermore, these diverse resources must be orchestrated to satisfy integrated yet diverse Quality-of-Service (QoS) requirements, which range from delay-sensitive control to bandwidth-hungry AIGC services.

Conventional network management approaches, including model-driven methods and Reinforcement Learning (RL), often struggle to navigate these complexities in large-scale, highly dynamic SAGIN. Model-driven approaches typically rely on idealized assumptions and static models, which often struggle to characterize the complex coupling among communication links, buffer queues, and computing nodes, leading to suboptimal performance in real deployments [2]. Conversely, while RL methods offer adaptability, they encounter inherent difficulties with scalability and sample efficiency when dealing with high-dimensional state-action spaces. More critically, these methods often operate as black boxes that lack the capability to interpret the semantic context of diverse QoS intents, such as distinguishing between emergency reliability and best-effort capacity needs. Given the increasing complexity of cross-layer interactions in SAGIN, there is an urgent need for more adaptive, knowledge-aware, and explainable intelligence in resource management [3].

Recent advances in large AI models (LAMs), particularly large language models (LLMs), have unlocked new paradigms for wireless resource optimization and network orchestration. Pre-trained LLMs exhibit strong capabilities in reasoning, in-context learning, and decision-making, and have been widely applied to diverse network management tasks [4]. While Retrieval-Augmented Generation (RAG) further enhances decision quality by grounding LLM outputs in real-time domain knowledge [5], most existing approaches remain limited to open-loop operations. Specifically, they predominantly deploy LLMs as single-shot optimizers handling static network snapshots, generating one-off strategies without continuous adaptation. Consequently, these solutions lack real-time state monitoring and execution feedback mechanisms, leaving them vulnerable to hallucinations and limiting their efficacy in dynamic, closed-loop resource management [6]. This progression, and the critical gap aiming toward closed-loop autonomy, is illustrated in the evolutionary timeline of Fig. 1.

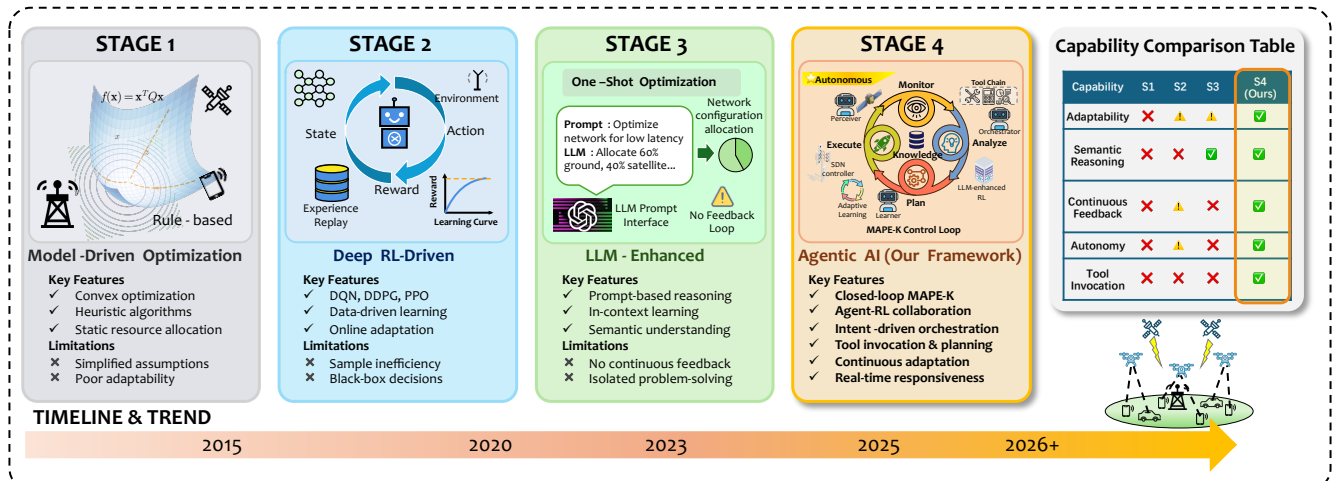

**Figure 1.** Evolution of AI-driven network management approaches for SAGIN, highlighting the progression from static model-driven optimization to autonomous agentic AI systems.

To achieve truly autonomous SAGIN resource management, we argue that it is necessary to elevate large models into agentic AI controllers. Agentic AI augments LLMs with capabilities such as tool invocation, memory management, and planning [7]. This enhancement facilitates a paradigm shift toward intelligent orchestration, where agents continuously perceive network conditions with semantic awareness. Specifically, they can interpret the contextual value of constrained resources, such as prioritizing satellite energy preservation over latency reduction, and decompose high-level operator intents into actionable tasks. Furthermore, these agents coordinate heterogeneous modules, including traditional algorithms and RL-based units, to iteratively refine strategies based on execution feedback. By effectively bridging semantic understanding with network-level operations via software-defined networking (SDN) interfaces, this paradigm is particularly well-suited for handling the dynamic and heterogeneous nature of SAGIN resource management [8].

In this paper, we present an agentic AI-driven framework for autonomous SAGIN resource management and make the following contributions:

• **Agentic AI control plane architecture:** We propose a closed-loop control framework built upon the Monitor-Analyze-Plan-Execute-Knowledge (MAPE-K) loop, incorporating three specialized AI agents for semantic resource perception, intent-driven orchestration, and adaptive learning. This architecture bridges the gap between operator intents and network-level execution through natural language reasoning and tool-augmented coordination, enabling autonomous adaptation to dynamic network conditions.

•**Hierarchical agent-RL collaboration mechanism:** We design a collaborative scheme where LLM-based orchestrators dynamically shape reward functions for RL agents based on semantic network conditions, facilitating policy adaptation to evolving environments. This hierarchical approach leverages LLMs for high-level semantic reasoning while delegating low-level real-time optimization to specialized RL algorithms, effectively combining interpretability with computational efficiency.

• **Validation through simulation:** We conduct a case study on task placement and resource allocation for AIGC services in energy-constrained SAGIN, benchmarking the proposed agent-enhanced RL against conventional baselines. Simulation results demonstrate a superior energy-latency balance and faster convergence, confirming the effectiveness of agentic AI for SAGIN resource orchestration.

## Why Agentic AI for SAGIN Resource Management?

SAGIN integrates resource-constrained non-terrestrial nodes with heterogeneous terrestrial networks, creating a large-scale, time-varying infrastructure where cross-layer resource awareness and real-time optimization are tightly coupled. However, conventional network management approaches struggle to deliver effective orchestration because of three inherent challenges.

•**Resource coupling and fragmented visibility:** The deep resource coupling between broad-coverage yet resource-limited non-terrestrial nodes and terrestrial networks fragments cross-layer visibility, hindering the acquisition of a unified global resource view. This necessitates semantic awareness to transform raw telemetry and network states into actionable cross-layer abstractions.

• **Performance heterogeneity and spatiotemporal dynamics:** Diverse nodes across space, air, and ground layers exhibit divergent physical attributes and transmission capabilities. These disparities are compounded by platform mobility, creating spatiotemporal dynamics that render static or segmented control mechanisms ineffective. Addressing this requires intelligent orchestration to dynamically coordinate cross-layer actions.

• **Conflicting QoS and decision complexity:** Diverse services ranging from latency-critical control to computation-intensive AIGC impose conflicting QoS requirements. This expands the decision space and amplifies uncertainty in joint offloading, caching, and routing, motivating adaptive optimization mechanisms that continuously refine policies based on execution feedback.

Together, these complexities justify the deployment of an agentic AI control plane that unifies semantic awareness, intent-driven orchestration, and adaptive optimization for autonomous resource management in dynamic SAGIN environments [9].

## Agentic AI Control Plane Architecture

To address these challenges, we propose an agentic AI control plane that functionally decomposes autonomous resource management into three core capabilities, namely semantic cross-layer awareness, intent-driven orchestration, and continuous policy adaptation. As illustrated in Fig. 2, these capabilities are realized through three specialized agent roles that collaborate to form a unified control plane for SAGIN resources. Rather than relying on a monolithic controller, this functional decomposition enables each agent to specialize in a distinct capability while maintaining coordinated operation through structured interactions [10]. Specifically, the control plane comprises the following three agent roles.

• **Semantic resource perceivers:** These agents bridge the gap between raw data and decision-making by fusing telemetry from satellites, aerial platforms, and ground nodes into unified, high-level descriptions of the global network state. By combining raw metrics, such as link quality, buffer occupancy, and energy levels, with domain knowledge from standards and historical logs, they create cross-layer views that highlight resource hotspots and bottlenecks in a format interpretable for both downstream algorithms and human operators. This directly addresses the issues of limited visibility and weak semantic abstraction.

• **Intent-driven resource orchestrators:** Acting as central planners, the orchestrators connect high-level objectives with concrete control actions. Given operator or service intents, such as minimizing delay under stringent energy budgets, they derive coordinated decisions on routing, spectrum allocation, and task offloading. Leveraging prior optimization cases and external knowledge, these orchestrators handle trade-offs among latency, reliability, and load balancing. They can also invoke specialized tools, including RL-based modules, classical optimization solvers, and SDN or orchestration APIs.

• **Adaptive learners and policy refiners:** By continuously monitoring key performance indicators (KPIs) and execution feedback, this agent role closes the loop between decisions and outcomes, enabling early detection of anomalies and performance drifts. Based on this feedback, they update internal memory and refine the prompts and reward functions used by other agents. This mechanism improves the robustness of policies under evolving conditions and provides intrinsic support for explanation and auditing, as the reasoning behind each update can be recorded and inspected.

To interact with the SAGIN infrastructure, these agents leverage a model context protocol (MCP) that provides unified interfaces to heterogeneous data sources and control tools [10]. Through MCP, agents can access telemetry databases for network monitoring, retrieve domain knowledge from standards documents and historical logs, invoke optimization solvers and RL-based modules for decision-making, and issue control commands to SDN controllers and orchestration platforms for strategy execution. While these agent roles establish the functional foundation for autonomous resource management, their effective operation requires a structured control workflow that coordinates monitoring, analysis, planning, and execution activities into a closed loop.

## Closed-loop Control via MAPE-K Mechanism

### A. MAPE-K Integration and Workflow

Building on the three specialized agents introduced above, we now describe their integration into a MAPE-K control loop to realize closed-loop resource management for SAGIN. The MAPE-K loop establishes a continuous cycle where network state drives decision-making and execution feedback enables adaptation. In this cycle, the Monitor phase collects multi-source telemetry and performance indicators from the SAGIN infrastructure. The Analyze phase then processes these observations into semantic resource views while evaluating their implications for resource management. Based on these insights, the Plan phase generates candidate cross-layer management strategies that reflect high-level intents and current network states. The Execute phase subsequently enforces selected strategies through SDN and orchestration interfaces. Throughout this process, the shared Knowledge component provides domain expertise, historical logs, and operational memory to all phases, supporting informed decision-making. The three agent roles integrate naturally into this loop to enable autonomous, adaptive resource management.

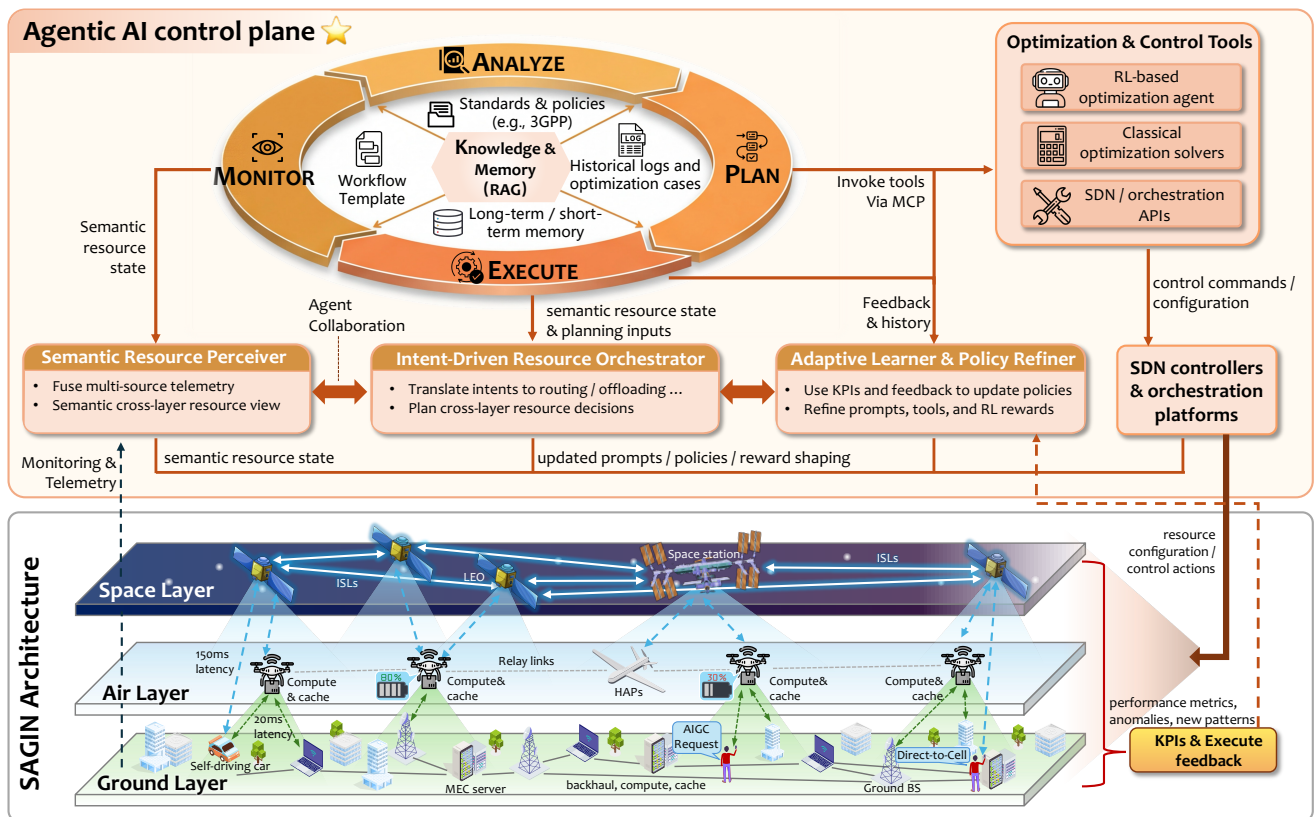

**Figure 2.** Agentic AI control plane architecture built upon the MAPE-K framework, incorporating semantic resource perceivers, intent-driven orchestrators, and adaptive learners to manage heterogeneous SAGIN resources.

### B. Monitor and Analyze: Semantic Resource Awareness

The Monitor phase continuously collects real-time, multi-dimensional network state data from the SAGIN infrastructure, encompassing traffic patterns, resource utilization, topology dynamics, and system anomalies across space, air, and ground segments. Distributed sensors and network probes gather operational data, which is then normalized for analysis. The Analyze phase employs semantic resource perceivers to perform comprehensive data fusion and state evaluation. Perceivers leverage advanced reasoning capabilities to process complex network patterns, integrating real-time observations with historical data and domain knowledge retrieved from the Knowledge base. This phase transforms raw telemetry into semantic cross-layer resource states that capture high-level network conditions, such as identifying relay link bottlenecks or energy constraints, thereby generating interpretable, decision-oriented views for the planning stage [11].

### C. Plan and Execute: Intent-Driven Optimization

Based on semantic resource states from the Analyze phase, the Plan phase formulates cross-layer management strategies such as spectrum allocation, routing adjustments, and task offloading. Intent-driven resource orchestrators generate action plans that balance competing metrics such as latency, reliability, and energy consumption [12]. Meanwhile, orchestrators query the knowledge base to select appropriate optimization tools for each specific task.

The Execute phase translates planning strategies into concrete network operations, distributing configurations to network segments through standardized control plane interfaces such as SDN controllers and orchestration APIs. Execution status and performance feedback are returned to adaptive learners, which refine internal memory and update reward functions for RL-based agents. This feedback mechanism enables continuous adaptation to evolving network conditions.

### D. Knowledge Management and Continuous Learning

The Knowledge component maintains two categories of information to support agent decision-making. Static domain knowledge includes wireless communication standards, network protocols, and resource management best practices that perceivers and orchestrators leverage to ensure compliance and informed reasoning. Dynamic operational knowledge comprises historical performance logs for few-shot learning, reusable workflow templates and skills for common management tasks, and interface documentation for tool invocation. Through RAG technology, agents query this knowledge base to retrieve contextually relevant information based on current network environment and optimization objectives. As the MAPE-K loop operates, execution feedback continuously enriches this knowledge base with new data and successful strategies, enabling progressive improvement of resource management capabilities.

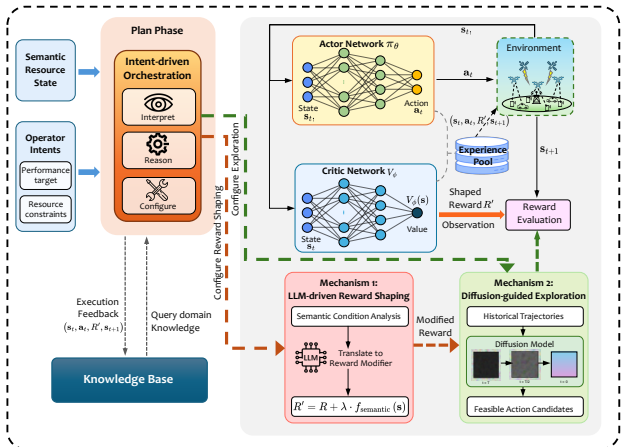

**Figure 3.** Hierarchical agent-RL collaboration in the Plan phase. The LLM-based orchestrator interprets operator intents and semantic states, then configures RL agents through dynamic reward shaping and diffusion-guided exploration.

## Agent-RL collaborative Optimization

While intent-driven orchestrators excel at semantic interpretation, tasks such as dynamic task offloading, real-time spectrum allocation, and adaptive routing in mobile SAGIN environments demand millisecond-level decisions over high-dimensional continuous action spaces. To address this challenge, the agentic AI framework adopts a hierarchical approach where orchestrators serve as high-level coordinators while invoking specialized RL agents for low-level optimization. Within the Plan phase, orchestrators interpret semantic resource states from perceivers, determine optimization objectives based on operator intents, and configure RL agents with appropriate parameters. RL agents then execute rapid decision-making, with execution feedback returned to adaptive learners for continuous policy refinement.

To enhance this collaboration, the framework incorporates two complementary mechanisms, as illustrated in Fig. 3. Orchestrators employ LLM-driven reward shaping to align RL optimization with evolving network semantics [13]. When perceivers report conditions such as satellite energy constraints or ground segment congestion, orchestrators generate modified reward functions. For example, under energy-constrained scenarios, the orchestrator adjusts rewards to penalize high-power transmissions while favoring energy-efficient routing paths. Meanwhile, the framework employs diffusion models to guide RL exploration. These generative models learn state-action distributions from historical trajectories stored in the knowledge base and generate exploratory actions that remain within feasible regions, substantially reducing ineffective exploration in high-dimensional optimization tasks [14]. These mechanisms enable seamless integration of semantic awareness while reactive optimization while ensuring RL-based decisions remain aligned with high-level network intents.

# Case Study: UAV-assisted AIGC Service Orchestration

## A. Scenario Description

We demonstrate the proposed framework through a UAV-assisted AIGC service orchestration scenario in SAGIN, as illustrated in Fig. 4. The physical layer depicts a LEO satellite constellation providing global coverage, a heterogeneous UAV cluster with varying energy states, and ground base stations with edge computing servers serving mobile users requesting real-time HD video generation services [15]. At the captured time instant, UAV-1 experiences critical energy depletion (25% remaining) while UAV-2 maintains adequate energy (80%), creating a dynamic resource management challenge that requires intelligent decision-making.

The agentic AI control plane demonstrates the complete MAPE-K operational workflow. Semantic resource perceivers continuously collect multi-dimensional telemetry and infer the semantic state "UAV cluster energy-constrained with satellite backup available but high latency." The intent-driven orchestrator receives this semantic state along with operator intent to minimize service latency while ensuring UAV energy sustainability. It then invokes a pre-trained Deep Diffusion Deterministic Policy Gradient (D3PG) agent to determine task placement across heterogeneous nodes, with communication routing and resource allocation. Crucially, the orchestrator shapes the reward function based on current network semantics to penalize energy-depleting UAV assignments while favoring satellite-assisted or ground processing. As shown in the embedded deep dive panel, this LLM-driven reward shaping mechanism enables the RL agent to rapidly adapt its policy. Finally, the orchestrator translates the resulting action into concrete SDN flow rules, resource allocations, and power control commands, with execution feedback returned to adaptive learners for continuous policy refinement. For instance, when the observed service latency exceeds the target threshold, the adaptive learner records this deviation and refines the orchestrator's reward configuration for subsequent decision cycles.

## B. Performance Evaluation

To validate both the effectiveness of D3PG algorithm and LLM-driven reward shaping, we conduct simulation experiments focusing on the task placement problem within the Plan phase. The D3PG agent determines AIGC task allocation across heterogeneous processing nodes and transmission routing selections to minimize service latency while managing UAV energy consumption. The reward function combines latency and UAV energy consumption through a penalty coefficient, which is dynamically shaped by the LLM orchestrator based on semantic network conditions. Given task placement and routing decisions, resources such as communication bandwidth, computation capacity, and transmit power are optimized through lightweight solvers to satisfy capacity and QoS constraints.

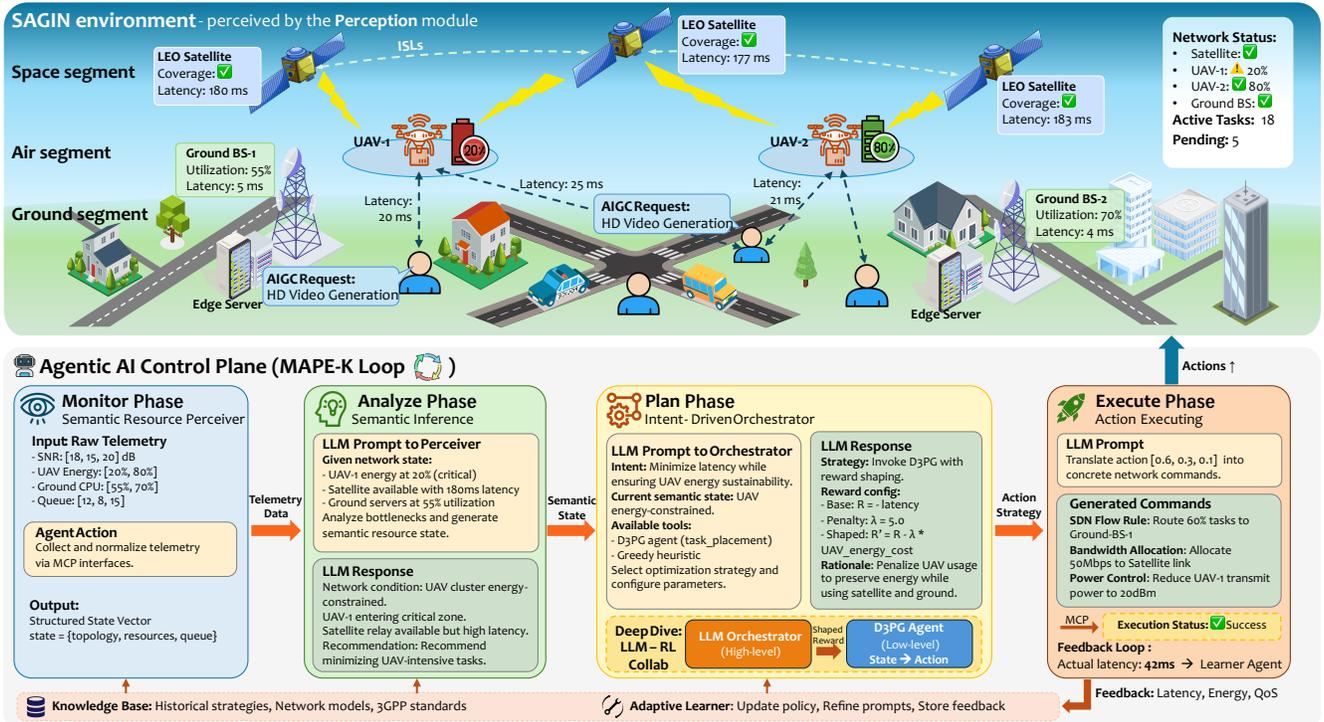

**Figure 4.** UAV-assisted AIGC service orchestration scenario demonstrating the MAPE-K workflow. The control plane perceives UAV energy constraints, adapts D3PG reward functions through LLM shaping, and executes optimized task placement decisions.

We implement the scenario using a simulator with 3 LEO satellites, 5 UAVs with varying energy states, 2 ground base stations, and 50 concurrent AIGC tasks under energy-constrained operational conditions. We compare five approaches to evaluate both algorithm effectiveness and reward shaping benefits: (1) LLM-shaped D3PG where the penalty coefficient is dynamically adjusted by the orchestrator based on UAV energy semantics, (2) Fixed-reward D3PG with constant penalty coefficient, (3) DDPG with LLM-shaped reward, (4) DQN with LLM-shaped reward, and (5) Greedy heuristic that assigns tasks to lowest-latency nodes. The RL agents are implemented in Python using PyTorch and trained over 1000 episodes with experience replay.

Fig. 5 presents the training convergence under energy-constrained conditions. The proposed LLM-shaped D3PG achieves the fastest convergence and highest episode reward, confirming that semantically driven reward shaping provides more informative learning signals than fixed reward functions. Fig. 6 further compares the converged service latency and UAV energy consumption. LLM-shaped D3PG attains the lowest average latency while reducing normalized energy consumption by 14% relative to the fixed-reward variant. This improvement stems from the adaptive penalty coefficient, which guides the agent to balance UAV utilization against energy depletion proactively. Among the baselines, DDPG and DQN with LLM shaping exhibit intermediate performance, confirming the advantage of D3PG's diffusion-based exploration in high-dimensional continuous action spaces, while the Greedy heuristic incurs substantially higher latency and energy cost due to its inability to plan beyond the current time step.

These results highlight a core design principle whereby the LLM orchestrator shapes the reward landscape on a slow timescale while the RL agent executes fast-timescale decisions, combining semantic reasoning with real-time responsiveness. This hierarchical agent-RL pattern is expected to generalize to other SAGIN tasks such as spectrum allocation and routing by replacing the semantic extraction rules and RL algorithm.

## Discussion and Future Directions

While this paper demonstrates the viability of agentic AI for SAGIN resource management, several promising directions deserve further exploration. Integrating RAG could enable dynamic knowledge acquisition from evolving technical standards and operational logs, reducing reliance on manually curated prompt templates. Extending the framework to multi-objective optimization scenarios, such as spectrum efficiency, fairness, and network resilience, requires advanced preference elicitation mechanisms and multi-agent coordination protocols. Addressing partial observability and adversarial network conditions through diffusion-based state estimation represents another promising research direction. Finally, transitioning from simulation to real-world deployment requires rigorous safety assurances, including formal verification of LLM-generated reward functions, explainability of RL-based decisions, and fail-safe mechanisms to handle ambiguous or conflicting semantic guidance.

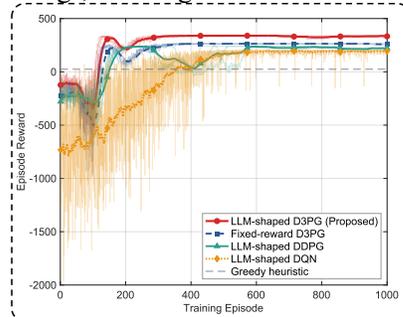

*Figure 5. Training convergence comparison across methods.*

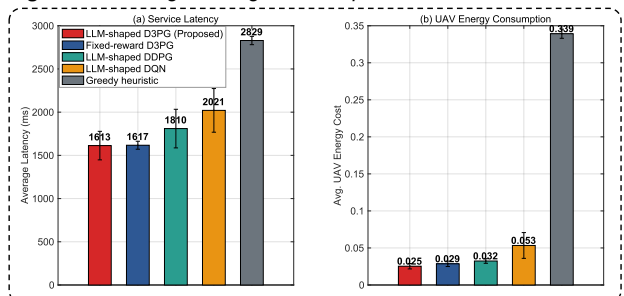

*Figure 6. Performance comparison across methods: (a) average service latency and (b) UAV energy consumption.*

## Conclusion

This paper presented an agentic AI framework for autonomous SAGIN resource management, establishing a control plane architecture built upon the MAPE-K loop with three collaborative AI agents, semantic resource perceivers, intent-driven orchestrators, and adaptive learners. The framework bridges high-level operator intents and low-level network optimization through hierarchical agent-RL collaboration, where LLM-based orchestrators provide semantic reasoning and goal-oriented guidance while specialized RL agents execute real-time decisions. Validation through UAV-assisted AIGC service scenarios confirms the framework's effectiveness in balancing multiple objectives under dynamic constraints. This agentic paradigm offers a generalizable approach to a broad range of SAGIN resource management problems, including but not limited to spectrum allocation, task offloading, and routing optimization, paving the way toward adaptive, AI-native SAGIN systems that autonomously interpret operator intents and respond to dynamic environments.